\documentclass[fleqn,usenatbib]{mnras}
\usepackage{amssymb}
\usepackage{newtxtext,newtxmath}
\usepackage[T1]{fontenc}

\DeclareRobustCommand{\VAN}[3]{#2}
\let\VANthebibliography\thebibliography
\def\thebibliography{\DeclareRobustCommand{\VAN}[3]{##3}\VANthebibliography}

\newcommand{\Mpc}{\ensuremath{\rm{Mpc}}}
\newcommand{\Smass}{M$_{\odot}$ }
\newcommand{\h}{\texttt{H1 }}
\newcommand{\myl}{\texttt{L1 }}
\newcommand{\f}{\ensuremath{f_0}}
\newcommand{\q}{\ensuremath{{q}}}
\newcommand{\pycbc}{\texttt{PyCBC}}

\usepackage{graphicx}	
\usepackage{amsmath}		
\usepackage{xcolor}   
\usepackage{color}
\usepackage{orcidlink}
\usepackage{subcaption}
\usepackage[normalem]{ulem}
\usepackage{array}

\title[SiGMa-Net II]{SiGMa-Net II: Distinguishing Binary Black Holes from Glitches}

\author[Narayan et al.]{Soorya Narayan\orcidlink{0009-0009-8338-3388}$^{1,2}$,\thanks{E-mail: soorya.science@gmail.com}
Anupreeta More\orcidlink{0000-0001-7714-7076}$^{3,4},$
Sunil Choudhary\orcidlink{0000-0003-0949-7298}$^{5},$
Sudhagar Suyamprakasam\orcidlink{0000-0001-8578-4665}$^{6}$
\newauthor{and Sukanta Bose\orcidlink{0000-0002-4151-1347}$^{3,7}$}
\\
$1$ Indian Institute of Science Education and Research, Pune 411 008, India\\
$2$ Aix Marseille Univ, CNRS/IN2P3, CPPM, Marseille, France\\
$3$ Inter-University Centre for Astronomy and Astrophysics, Post  Bag 4, Ganeshkhind, Pune 411 007, India\\
$4$ Kavli Institute for the Physics and Mathematics of the Universe (IPMU), 5-1-5 Kashiwanoha, Kashiwa-shi, Chiba 277-8583, Japan\\
$5$ Department of Physics, University of Western Australia, Crawley WA 6009, Australia\\
$6$ Nicolaus Copernicus Astronomical Center, Polish Academy of Sciences, Bartycka 18, 00-716, Warsaw, Poland\\
$7$ Department of Physics \& Astronomy, Washington State University, 1245 Webster, Pullman, WA 99164-2814, USA\\
}

\date{Accepted XXX. Received YYY; in original form ZZZ}

\pubyear{2024}
\begin{document}
\label{firstpage}
\pagerange{\pageref{firstpage}--\pageref{lastpage}}
\maketitle

\begin{abstract}
With increasing sensitivity of the gravitational wave (GW) detectors, we expect a significant rise in the detectable GW events. To process, analyse and identify such large amounts of GW signals arising from mergers of Binary Black Holes (BBH), we need both speed and accuracy. In the search for (massive) BBH signals, the biggest hurdle is posed by the various non-gaussian noise transients called glitches. Compared to our previous work, which used a simple convolutional neural network to distinguish BBHs from Blip glitches, this work uses transfer learning with InceptionNetV3 to distinguish BBHs from six types of most popular glitches from the third observing run of LIGO. While the glitches are real and identified via GravitySpy, the BBH signals are simulated and then injected into the real detector noise for each of the two LIGO detectors. We generate Sine-Gaussian Projection (SGP) maps by cross-correlating data with Sine-Gaussian functions of varied quality factors ($Q$) and central frequencies ($f_0$) and projected on the $Q-f_0$ plane. We find that SGP maps make it easier to distinguish BBHs from glitches that look very similar to BBHs in the Time-Frequency maps like the Blips, while also maintaining significant morphological differences between BBHs and the more frequent glitches - Scattered Light and Fast Scattering. Our network has an accuracy of 87\%, a TPR of 0.83 for an FPR of 0.1 on our test dataset. It is also robust, retaining its level of accuracy, when tested on real BBH events identified in the first three observing runs of LIGO. Our proposed method shows the viability of using the SGP maps and neural networks for fast identification of GW events improving the efficiency of standard search pipelines.
\end{abstract}


\section{Introduction}
Over the last decade, gravitational waves has opened new avenues to study the Universe. The Advanced Laser Interferometric Gravitational-Wave Observatory (LIGO) \citep{Aasi_2015}, the Advanced Virgo \citep{Acernese_2015}, and the recent addition, KAGRA \citep{2019}, have observed gravitational wave signals from merging Binary Black Holes (BBH), Binary Neutron Stars (BNS), and Blackhole-Neutron Star (BNS) systems. 
Over the three observing runs (O1 to O3) of LIGO--VIRGO--KAGRA, the number of confirmed gravitational wave events has increased from 3 events in O1 to 8 events in O2 to 79 events in O3 \citep{PhysRevX.9.031040, PhysRevX.11.021053, 2024PhRvD.109b2001A, PhysRevD.101.083030, PhysRevX.13.041039, Nitz_2020} to 128 events in O4a \citep{2025arXiv250818082T}. The reason for such a drastic increase from one run to the next is two-fold. 

The primary reason is an increase in the sensitivities of each detector. With increasing sensitivities, the detectors can pick up on faint signals that would have otherwise gone unobserved. The flip side of this increased sensitivity is that the detectors now pick up even more noise. This includes typical Gaussian noise observed everywhere and the more problematic, non-Gaussian, short-duration transients called glitches \citep{Blackburn_2008}. Glitches deteriorate the quality of GW merger searches. The non-Gaussian nature of glitches increases the False Alarm Rates (FAR) of these searches, making it harder to ascertain the true nature of a trigger \citep{Abbott_2020}. They have made it especially difficult to confidently identify CBC signals sourced by a system with a high total mass \citep{Abbott_2018, Nitz_2018}.

\begin{figure*}
    \centering
    \includegraphics[width=\linewidth]{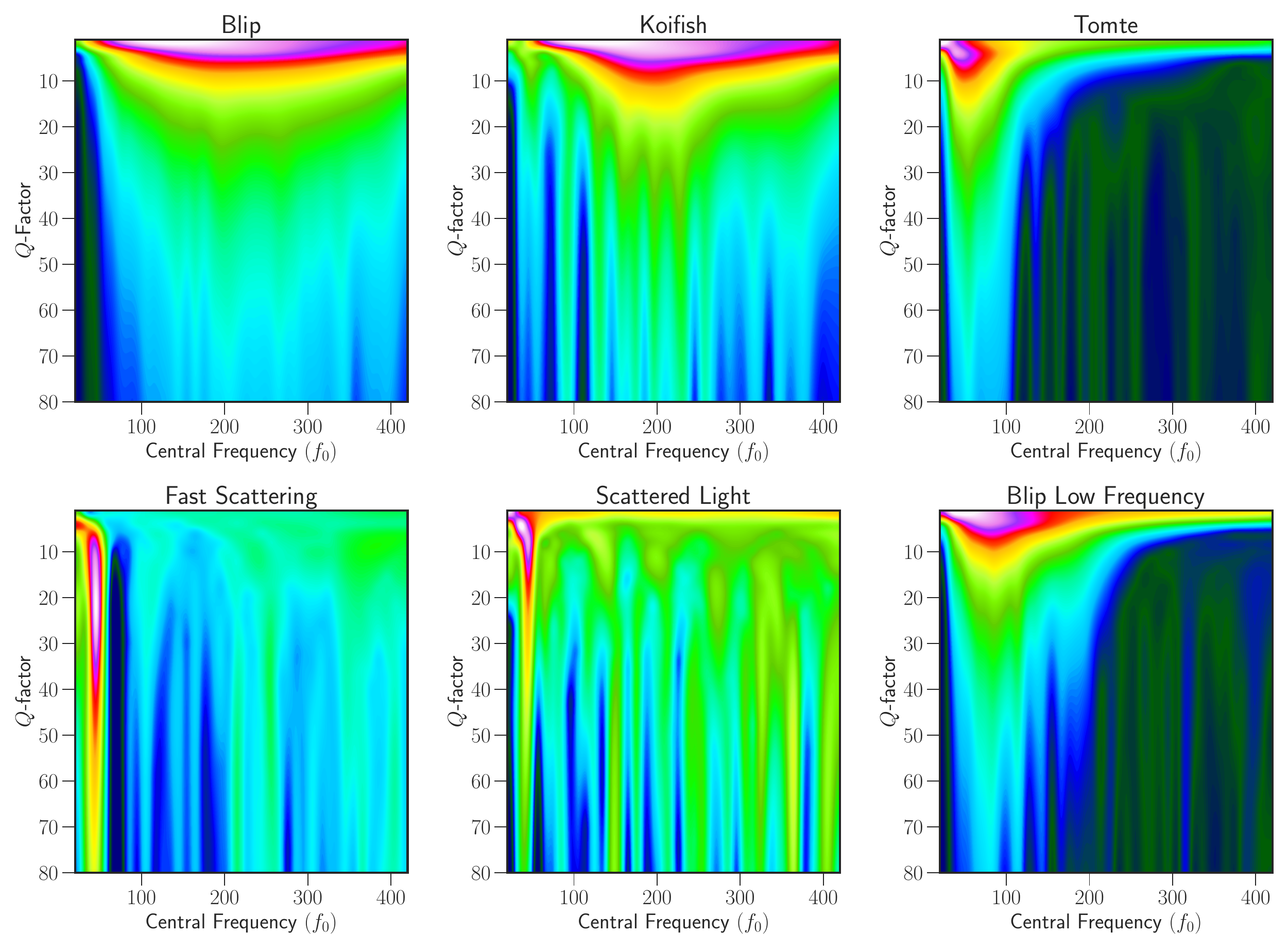}
    \caption{Sine Gaussian Projections of the siz different types of glitches used to train the network. The Blip and Koifish glitches span the entire frequency axis of the space, while the other glitches are localized to the low-frequency region of the space. Fast scattering glitches extend along the Q-factor axis while the other glitches typically stay in the small Q-factor region.}
    \label{fig:ref_glitches}
\end{figure*}

GravitySpy \citep{Zevin_2017} is the current state-of-the-art for classifying glitches. There are a total of 24 types of glitches, including the ``None of the Above'' category for all the glitches that do not fall into the other categories \citep{Glanzer_2023}. These classes are identified based on the glitch morphologies in Time-Frequency maps. In O3 alone, GravitySpy \citep{Zevin_2017} identified  $\sim 100,000$ glitches in the \h detector and $\sim 180,000$ glitches in the \myl detector with a confidence of $\> 95\%$ \citep{Glanzer_2023}. The sheer difference in the order of magnitude of the number of glitches to the number of signals demands faster and better data analysis techniques to discriminate between glitches and signals. Additionally, the similarity of some glitches to BBH merger signals necessitates the development of complex denoising and discriminating algorithms. 

Another reason for the increasing number of events identified in the later runs is the increase in the duration of each observational run. For reference, O1 lasted for less than half a year, O2 lasted for approximately one year, O3 lasted for slightly over a year, and O4 lasted a little over two years \citep{ligo_observing_plan}.

The increase in the duration of each observational run also leads to an overall increase in the amount of data to be processed. This will be further amplified when future ground-based detectors like LIGO-India \citep{Saleem_2022}, Einstein Telescope \citep{2020JCAP...03..050M}, Cosmic Explorer \citep{2019BAAS...51g..35R}, and Neutron Star Extreme Matter Observatory (NEMO) \citep{2020PASA...37...47A} join the network of detectors. Therefore, faster techniques like neural networks start to shine in comparison to traditional modelling and statistical techniques, for example, \cite{PhysRevD.103.044035, Davis_2020, Nitz_2018}. 

As a first step towards developing such techniques, \cite{PhysRevD.107.024030} proposed SiGMa-Net, a CNN model, to distinguish between BBH merger signals and Blip glitches. Traditional techniques like template matching are very accurate but take very long for each sample. SiGMa-Net matches, and sometimes surpasses, the accuracy of these techniques but only takes a fraction of the time. Along with the CNN, \citep{PhysRevD.107.024030} also introduced a new space to visualise gravitational wave time series, the Sine Gaussian Projection (SGP). SGP were introduced as an alternative to Time-Frequency maps where the Blip glitches take a similar morphology to that of BBH signals. Blip glitches \citep{Cabero_2019} are notorious for impacting the efficiency of CBC searches owing to their similarities to BBH signals \citep{Abbott_2020}. 

This work takes SiGMa-Net a step forward by expanding the goal to distinguishing BBHs from multiple types of glitches using SGP maps. \cite{Glanzer_2023} provides the statistics of each type of glitch in each of the detectors in the O3 run as classified by GravitySpy \citep{Zevin_2017}. In an attempt to work with the most frequent glitches in the detectors and those with similarities to BBH merger signals in the SGP maps and/or TF maps, we decided to use six types of glitches. In terms of frequency of observation, we chose Scattered Light and Fast Scattering. In terms of morphological similarities in the TF domain, we chose  Blips and Koifish. Lastly, in terms of the morphological similarities in the SG domain, we chose Blip Low Frequency and Tomte glitches. Figure \ref{fig:ref_glitches} shows the various glitches in the SG domain.

Another improvement in this work was decreasing the computational time for generating SGP maps. \cite{PhysRevD.107.024030} generates SGP maps for short-duration data chunks in $\sim 30s$, which is too long for up-scaling to this work and future, larger works.

In summary, this paper tackles two goals: first, to speed up the SGP map generation process, and second, to expand the SiGMa-Net to distinguish a larger number of glitches from BBH signals. This paper is arranged as follows. In section \ref{sec:methods}, we describe the data used for this work, how the samples are generated, and the network architecture used in this work. In section \ref{sec:results}, we talk about the network performance and the nuances in data selection that affect the network. In section \ref{sec:conclusions}, we discuss the prospects of SiGMa-Net and what direction to take the next steps in. 

\section{Methods} \label{sec:methods}
In this section, we describe the format of the input images, how they were prepared, and the details of the neural network.

\subsection{Sine-Gaussian Projection Maps}
We use Sine Gaussian Projection (SGP) maps instead of the more common Time-Frequency maps to project the time series data into a plane. The SGP space is represented by two variables, central frequency (\f) and Quality Factor \q. We fix the range of \f$\thinspace$ and \q$\thinspace$ domains such that all the samples can be visualised without having to change the domains. We sample points evenly from the selected domains to produce sine-Gaussian templates onto which the time series data is projected. We then plot the projection values corresponding to each pair of \f$\thinspace$ and \q$\thinspace$ as a color map where the pixel values are smoothed using spline interpolation. The colors denote the intensity of the projection. In this work, we use $\f$\thinspace$ \in [20, 430]$ and $\q \in [1,80]$. 

\begin{figure*}
    \centering
    \includegraphics[width=\textwidth]{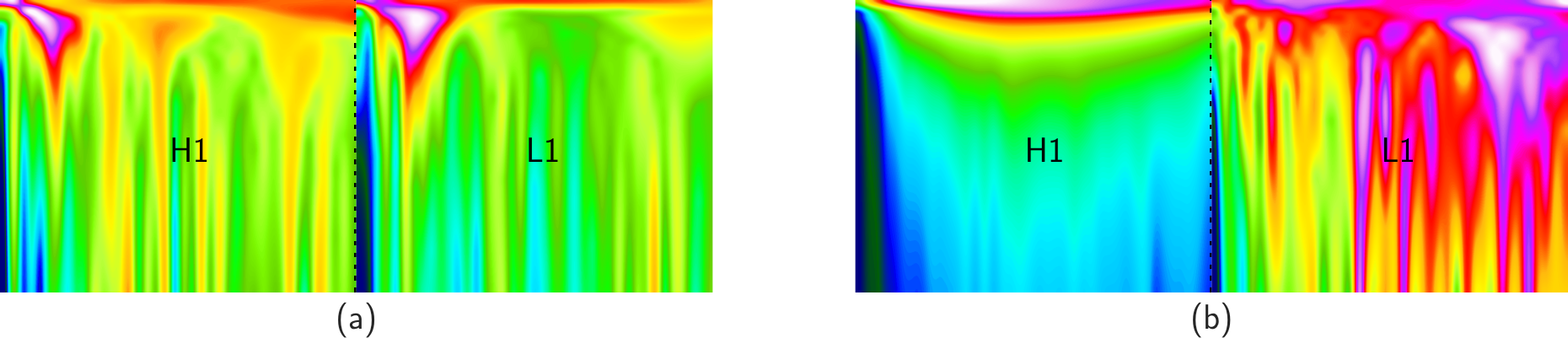}
    \caption{Example images used in SiGMa-Net. Panel \textbf{(a)} shows the event GW190706\_222641. Both detectors show similar projections, indicating the astrophysical nature of the source of the signal. The left half of the image shows the \h projection and the right half the \myl projection. Panel \textbf{(b)} shows a Blip sample used to train the network. The Blip glitch is detected in the \h detector, and the corresponding data from \myl shows random noise. The two halves are labelled as "H1" and "L1", and a dashed vertical line down the middle demarcates the boundary between the two projections.}
    \label{fig:network_input}
\end{figure*}

Figure~\ref{fig:network_input} shows two examples of the input to our network. In panel \textbf{(a)}, we show the SGP maps of the event GW190706\_222641 where strong projections are visible for both the \h and the \myl detectors. In panel \textbf{(b)}, we see the projection of a Blip glitch in the \h detector (left half) and random noise in the \myl detector (right half). The two halves are labelled as "H1" and "L1" with a vertical dashed line down the middle of each panel denoting the boundary between the two projections. The annotation is to help the reader distinguish the two halves. The network is trained on images without any such annotations.
BBH merger signals have an astrophysical origin and will most likely be observed in both detectors at close times. While terrestrial events like glitches will be observed primarily in one of the two detectors. The rate of observation of glitches simultaneously in both detectors can be estimated using time-shift techniques as the ones implemented/referenced in \cite{Abbott_2016, Abbott_2016b, Abadie_2012, Abbott_2019b, Abbott_2021}. A more detailed explanation for SGP maps and related choices and assumptions are given in \cite{PhysRevD.107.024030}.

\subsection{Datasets} \label{sec:data}
The construction of training sets that closely resemble real data is the first crucial step in the training of any neural network. It ensures the smooth translation of the neural network from a hand-picked training data set to real-world data. Although there are a lot of glitches in the real data \citep{Glanzer_2023}, the number of BBH merger signals is very small. Thus, we make use of simulated BBH signals, which are then injected in real noise extracted from the publicly available in O3 LIGO data. To provide accurate noise information to the network, we fetch both \h and \myl noise segments that do not overlap with eachother, and use them for corresponding halves of the image.

We fetch all of our data using the frame type: \texttt{\{ifo\}\_GWOSC\_\{RUN\}\_4KHZ\_R1} and channel name: \texttt{\{ifo\}:GWOSC-4KHZ\_R1\_STRAIN}. For real detector noise, used with our simulated BBH samples, we randomly pick days during O3a and check the daily plots showing glitch rates binned in time and SNR. We select the GPS times with the least number of glitches and on average, the least SNRs by visual inspection of the plots of glitch rates. Since these segments are not completely void of any artifacts, we manually go through Time-Frequency plots, Sine-Gaussian Projection plots, and Time Series plots of 4-second-long chunks of the data to identify artifacts. For the samples with artifacts, we remove 8s of data centered at the artifact. To further ensure randomness in the simulated BBH samples, we retrieve noise for \h and \myl with data from multiple days without any overlap. We use 103,000 seconds of data for \h noise and 83,000 seconds of data for \myl noise. In contrast, \cite{PhysRevD.107.024030} uses a single 64s noise file from \h for its entire dataset for both parts of the image. Trying the same in this work led to overfitting. 

The range and distribution of parameters for the simulated BBH sample are as shown in Table~\ref{bbh_params}. 
Noise files are selected randomly from the available cleaned noise files. These files can range from 12s to 1628s. We estimate the PSD from the entire duration of the noise file. From the chosen noise file, 2s of data is selected randomly and the BBH signal is superimposed. If the duration of the signal is greater than 2s, then the 2s around the largest amplitude in the signal is retrieved and combined with the noise chunk. If the signal is shorter than 2s, then the two are added such that both their centers coincide. This combined time series, along with the PSD, estimated from the entire noise file, is used to generate the SGP map. 

In short, we generate BBH merger signals, project them onto each detector using the \texttt{\pycbc} package's \texttt{project\_wave()} function, add real-O3 noise corresponding to each detector, and then join the two images side-by-side. 

We use the GravitySpy Catalog to find glitches. The types of glitches are selected based on their frequency in O3 \citep{Glanzer_2023} and their similarity to BBH signals in morphology in either TF space or SG space. The glitches resembling BBH signal morphology in TF space tend to have minimal resemblance in SG space. We select triggers that correspond to glitches that were classified with a confidence level of 60\% or above. We then fetch 16 seconds of data with the trigger at the center, i.e., from $[t_{\rm trig} - 8s, t_{\rm trig} + 8s]$, where $t_{\rm trig}$ is the GPS time of the glitch. Similarly, we fetch the 16s of data corresponding to the same GPS time in the second detector. If we find that for a particular trigger, the data is unavailable in the other detector, we remove said trigger from our dataset. Similarly, if only a part of the 16s is available in the detector, then we remove the trigger from our dataset.

Generating SGP maps for glitches and BBH signals is slightly different. For glitches, we use the time-series data of a particular trigger directly. We use the entire 16s segment to estimate the PSD. This PSD is then used to whiten the segment. Then we retrieve the 2~s of data from the center, which helps emphasize the short-lived glitch in our projection. The 2~s segment, along with the estimated PSD is used to generate the SGP map. A detailed description of how to produce an SGP map is provided in \cite{PhysRevD.107.024030}. Figure~ \ref{fig:ref_glitches} shows the various glitches in the Sine-Gaussian space.  

We prepare test sets of superevents identified in the \texttt{GstLAL, MBTAOnline, PyCBC} and \texttt{SPIIR} pipelines to test the network's performance. These samples will be referred to as triggers for the remainder of this article. The triggers were selected from The Gravitational-Wave Candidate Event Database (GraceDB) for dates between April 2019 and March 2020 (dates of O3). The event GPS time is constrained to remain within the DMT-OBS-INT flag, further, the GPS time should contain data from both \h and \myl detectors. We also put a maximum FAR limit of $10^{-6} {\rm Hz}$. We obtain a total of 9,566 samples coming from all four pipelines. To obtain labels for these samples, we cross-reference the GPS time to the GravitySpy catalog and the list of known BBH merger events. We find labels for $\sim 450$ samples. 

Apart from the above-mentioned triggers, we also prepare a dataset using BBH mergers from the first machine learning gravitational-wave search mock data challenge (MLGWSC-1) \citep{Schafer:2022dxv}. 
MLGWSC-1 compares four machine-learning techniques designed to identify BBH signals and two traditional, analytical techniques. MLGWSC-1 produces three datasets of BBH merger signals injected in simulated noise and one dataset with real noise from the O3a run.
Our MLGWSC-1 dataset uses dataset 4 from MLGWSC-1 where the BBH merger signals are injected in real O3a noise, and the six types of glitches used in this work. We use the scripts provided in \cite{https://doi.org/10.48550/arxiv.2209.11146} to generate the time series data for the BBH mergers injected in real O3 data. We also produce a list of gps times where the BBH signal was injected. We then use the time series to generate the SGP maps following the procedure we followed for glitches. We take 16s of data centered at the injection time and generate SGP maps. We produce 6000 BBH samples for the MLGWSC-1 dataset used in this work. We combine the BBH merger samples with 6000 glitches (1000 of each category) to produce the final MLGWSC-1 dataset.

\begin{table}
\begin{center}
    \begin{tabular}{|c|c|c|}
    \hline\hline
        \multicolumn{3}{|c|}{\textbf{Simulated BBH Parameters}} \\
        \hline
        Parameters & Range & Distribution \\
        \hline
        Component Mass Range & [8,100] \Smass  & Uniform\\
        Inclination  & [0,$\pi$] & Sine\_angle \\
        Polarization &  $\left[-\frac{\pi}{2}, \frac{\pi}{2}\right]$ & Uniform\_angle \\
        Right Ascension + Declination & [0, $2\pi$] , $\left[-\frac{\pi}{2}, \frac{\pi}{2}\right]$ & Uniform\_sky \\
        Distance & [300,9000] $\Mpc$ & Uniform\_radius\\
        \hline\hline
    \end{tabular}
    \caption{\label{bbh_params} Parameters used for simulating BBH signals. 
    A total of 50,000 unique samples were generated.}
\end{center}
\end{table}

\subsection{Populating the Datasets}
Our training dataset consists of 48,000 samples, with 24,000 belonging to the BBH class and 24,000 belonging to six types of glitches (4000 for each glitch type). Our testing and validation datasets consist of 12,000 samples each, divided equally among the two classes and further divided equally among the 6 types of glitches. We randomly select the BBH samples from the 50,000 injections we initially produced. We also randomly select the glitch type-by-type from all the available samples. Each glitch amounts to one-sixth of the total number of glitches in a particular dataset.

\begin{figure*}
    \centering
    \includegraphics[width=\linewidth]{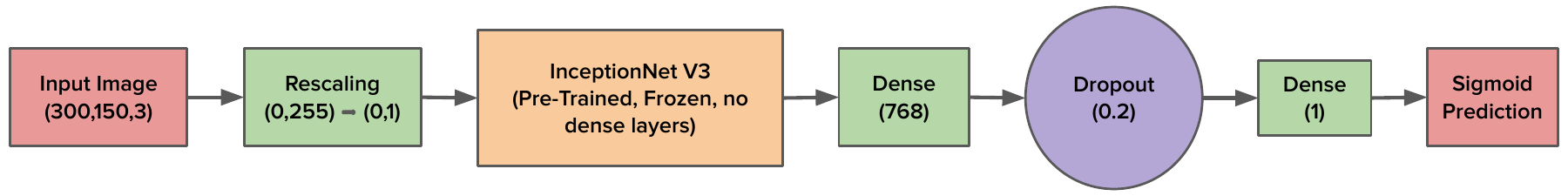}
    \caption{The network architecture of SiGMa-Net. The pre-trained weights are frozen, and the final dense layers are replaced with custom, trainable dense layers. }
    \label{fig:network_diagram}
\end{figure*}

\subsection{The Network} \label{sec:Network}
We use transfer learning to produce a network that can distinguish between BBH merger signals and six different types of glitches.
The motivation for using transfer learning comes not from the lack of availability of labeled datasets but from the fact that these pre-trained models prove to be very robust even after re-training \citep{yosinski2014transferablefeaturesdeepneural, han2021robusttransferlearningpretrained, kornblith2019betterimagenetmodelstransfer}. 

Our network was able to correctly identify a good fraction of the BBH samples from a set of $\sim 80$ BBH samples after re-training as you will see in Section \ref{sec:results}

We use the pre-trained InceptionNetV3 model in this work. InceptionNetV3 is the latest variant of the InceptionNet published in \cite{2014arXiv1409.4842S}. The most obvious (as you can see in this work as well) way to improve the performance of deep neural networks for any computer vision task would be to increase the size of the network. By increasing the number of layers and the size of each layer, one can produce networks of higher quality. However, this technique requires the availability of large labeled datasets. Aside from the requirement for large labeled datasets, there are two major drawbacks: the networks will be prone to overfitting, especially if the labeled dataset is limited, and the networks also demand significantly more computational resources to train. The conception of the inception network was the result of solving the two drawbacks with the use of sparsely connected architectures \citep{2014arXiv1409.4842S}. 

Inception networks were designed to produce deep representations at each layer. Inception networks use multiple kernel sizes at each layer and then concatenate the results before feeding them into the next layer. This results in a wider network and a deeper representation at each layer. Inception networks typically make use of $1\times 1$, $3\times 3$, and $5\times 5$ convolution kernels along with $3\times 3$ maxpool layers. The architecture proposed by \cite{2014arXiv1409.4842S} passes the output of one layer into a $1\times 1$ convolution before feeding the $3\times 3$ and $5\times 5$. They also add a $1\times 1$ convolution between the $3\times 3$ max pool layer and concatenation. The $1\times1$ convolutions allow for retaining unique features while reducing computation costs. Later iterations of inception networks made use of asymmetric kernels like $1\times 5$, $1\times 7$, $5\times 1$ etc., in parallel in place of large kernels like the $5\times 5$ and $7\times 7$ to further reduce the computation whilst avoiding a representational bottleneck \citep{szegedy2015rethinkinginceptionarchitecturecomputer}.

InceptionNetV3 implements factorised asymmetric convolutions up to a kernel size of $1\times 7$. The asymmetric convolutions are implemented in parallel, avoiding a representation bottleneck and thus making the network wider instead of deeper. This results in much shorter training times with very little sacrifice in performance \citep{szegedy2015rethinkinginceptionarchitecturecomputer}

The input images are of size $300\times 150\times 3$, where 3 refers to the RGB channels of the image. The pixel values are rescaled to $[0,1]$ before being fed into the network. We discard the final dense layers of the pre-trained InceptionNetV3. We freeze the weights of the remaining layers for training. We flatten the output of the final convolutional block through a 2D Global Average Pool (\texttt{GlobalAveragePooling2D()} in \texttt{Keras}; hereafter referred to as \texttt{GAP}). \texttt{GAP} is a downsampling technique typically used in conjunction with convolutional networks to reduce the spatial dimensions while preserving channel-wise information. We then feed the output of the \texttt{GAP} into a dense layer with 768 nodes, followed by a dropout layer with a dropout probability of 0.2, the output of which is passed into the prediction layer with a sigmoid activation function. Our BBH class is set to class 0, and our glitch class is set to class 1.

\section{Speeding Up SGP Generation Process}
In \pycbc, we perform matched filtering between a bank of BBH templates and a data segment containing the signal. We can determine an SNR based on how well the data segment matches each BBH template. We produce a sine-gaussian template in the frequency domain for a pair of \f$\thinspace$ and \q$\thinspace$ values. Our SGP maps comprise a $40\times40$ grid of \f$\thinspace$ and \q$\thinspace$ showing the SNRs obtained when data is matched with the sine-Gaussian templates. The matched filtering takes one template at a time, and since our SGP map has 1,600 phase plane points, we need to run the operation 1,600 times. This was done sequentially in \cite{PhysRevD.107.024030} and took $\sim 30$~s per SGP map. To accelerate our entire workflow, our first step was reducing the computation time to produce SGP maps. After trying various techniques along the lines of parallelization and vectorization, we decided to use GPUs for the task. This sped up the computation times by a factor of $\sim120$.

Since the relevant functions in \texttt{PyCBC} primarily use CPUs, we adapted them to work on GPUs. We made use of the \texttt{CuPy} package for this task. Since GPU memory is typically not as large as CPU memory, it puts a constraint on the size of the time series that can be converted into an SGP map at once. Another caveat is that if one were to use the GPU code to generate a single map, then the compilation (run once at the start of execution) time takes up a significant fraction of the total time, resulting in a speed increase by a factor of $\sim35$. But the subsequent images don't require the compilation process, so we see an average increase in speed by a factor of $120$. The computation times for various steps are summarised in Table~\ref{sgp_speed}. The original code was run on Intel Xeon Gold 6248 CPUs. The GPU code was run on a Nvidia Tesla K40m (12GB memory).
The increased speed allowed us to process large datasets to train our network.

\begin{table}
\begin{center}
    \begin{tabular}{|m{2cm}|m{2.5cm}|m{2.5cm}|}
        \hline\hline
        \multicolumn{3}{|c|}{\textbf{Code Version Comparison}} \\
        \hline
        Version & Time taken for generating a single map & Average time taken per map when generating 40 maps\\
        \hline
        Original Code & 35 s & 36 s \\
        GPU version & 0.9 s & 0.3 s \\
        \hline\hline
    \end{tabular}
    \caption{\label{sgp_speed} Speed comparison of the original CPU code and the revised GPU code. The original code was run on a single core of an Intel Xeon Gold 6248. The GPU code was run on a Nvidia Tesla K40m. The GPU code is $\sim \times 120$ faster when generating multiple maps, and $\sim \times 35$ faster for generating a single map.}
\end{center}
\end{table}

\section{Results} \label{sec:results}
In this section, we explain the various aspects of the network and its performance on diverse combination of BBH and glitch datasets. 
\subsection{Performance on various datasets}
\begin{figure}
    \centering
    \includegraphics[width=0.9\linewidth]{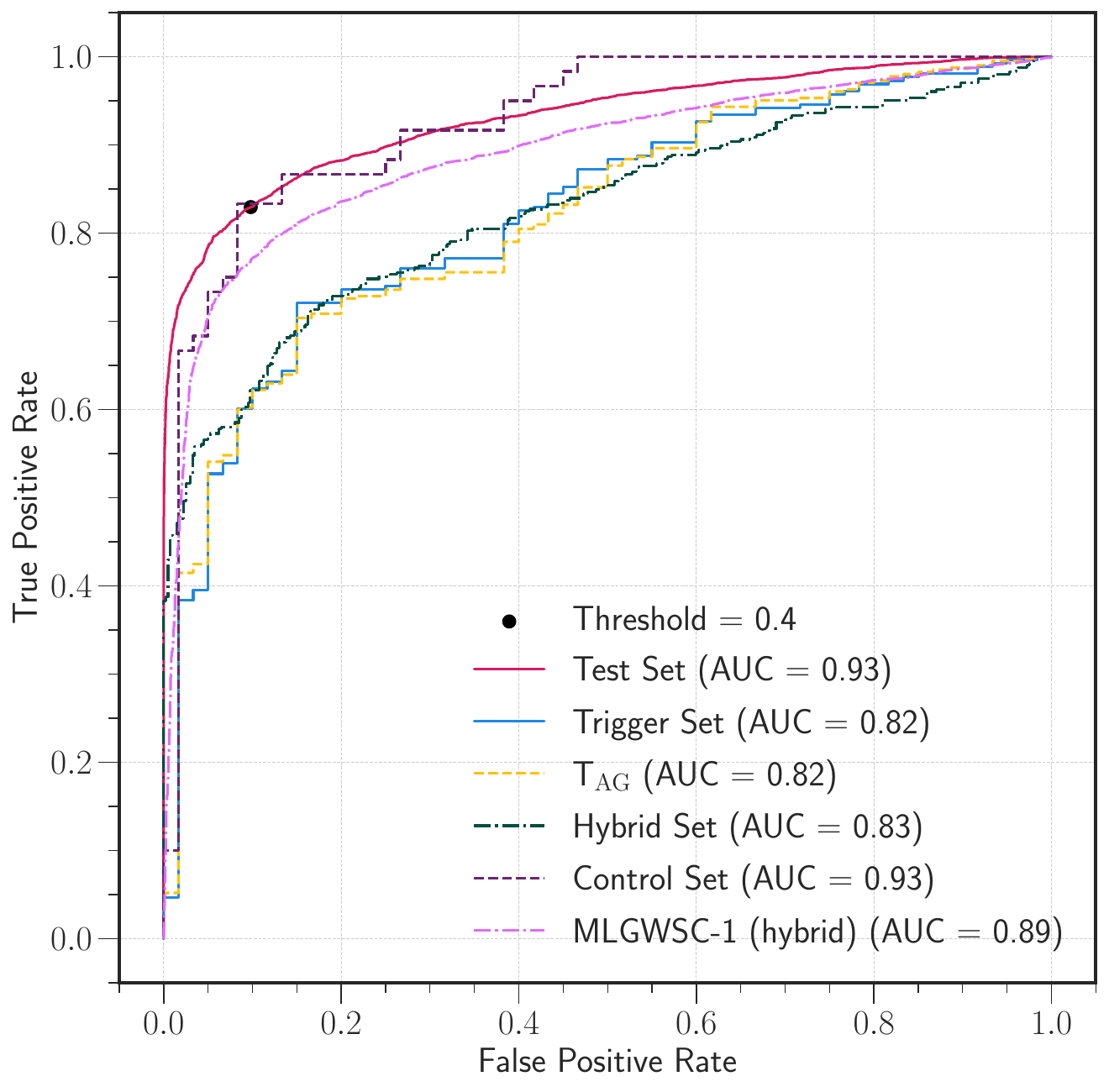}
    \caption{The ROC curves for various test datasets which are detailed in Section~\ref{sec:results}. The black dot shows the selected threshold for this work, $0.4$.The network performance is similar for the Test and Control datasets and are better than others with the former dataset comprising simulated BBHs and the latter real BBH events. }
    \label{fig:Network_ROCs}
\end{figure}

\begin{figure*}
    \centering
    \includegraphics[width=0.9\linewidth]{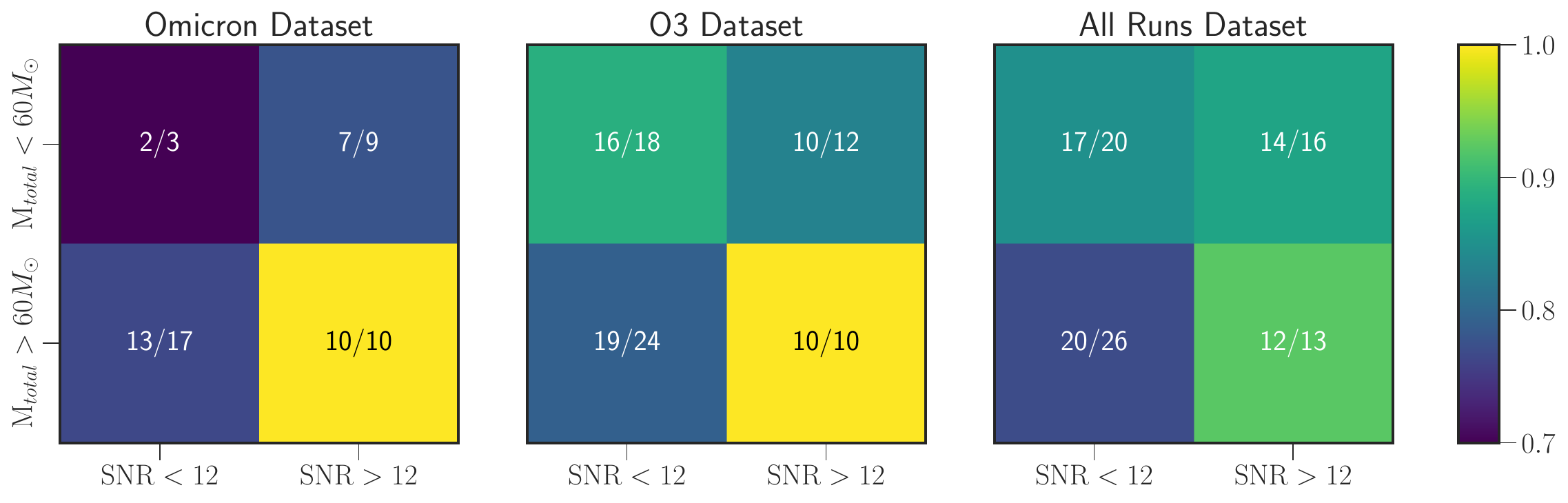}
    \caption{Performance of the network, for a threshold of 0.4,  when classifying real BBH events binned by their total masses and SNRs.The All Runs dataset contains all of the BBH events from the O1, O2 and O3 observing runs. The fractions indicate correctly classified events out of the total events in a given bin. The colors depict the accuracy of the network and seem to change slightly across the datasets, but only because of the low statistics. }
    \label{fig:Real_GW_performance}
\end{figure*}

We evaluate and quote the performance using standard metrics like the Receiver-Operating Characteristic (ROC) curve, the Area Under the ROC Curve (AUC), and F1 scores. An ROC curve plots the True Positive Rate (TPR) against the False Positive Rate (FPR) at various classification thresholds. This helps gauge the expected TPR and FPR for any selected threshold. The AUC is a single-value metric that quantifies the performance of the network on the entire test dataset. A perfect network will obtain an AUC of 1, while a random classifier will obtain a 0.5. The F1 score is the harmonic mean of the network's precision and recall. This is especially useful for evaluating a network on imbalanced datasets. 

Below, we describe the various test datasets used in this analysis.
The Test dataset consists of 6,000 simulated BBH samples drawn from the same distribution as the training dataset and 6,000 glitches, where each of the 1000 samples belongs to a glitch of a particular class.

The Trigger dataset consists of real BBH merger signals observed in LIGO (refer \ref{sec:data} for information on how these were selected). The glitch category contains Blips, Koifish, Tomte, Fast Scattering, Scattered Light, and Blip Low-Frequency samples identified from the \texttt{GstLAL, MBTAOnline, PyCBC} and \texttt{SPIIR} (see Section~\ref{sec:data}) search pipelines. Their labels are assigned based on the classification by GravitySPy \citep{glanzer_2021}. Glitches without corresponding labels are not considered. In total, we have 60 BBH merger signals and 258 glitch samples (all six glitch classes combined).

The T$_{\rm{AG}}$ dataset consists of the same real BBH merger signals as the Trigger dataset, but contains all 20 glitch classes and the \textit{None of the Above} that matched the available triggers from the aforementioned search pipelines. This increases the samples in the glitch category to 405. 

The Hybrid dataset has the simulated BBH samples from the Test dataset and the same glitches as those in the T$_{\rm{AG}}$ dataset.
Since we only have 405 glitch samples, we randomly select 400 simulated BBH samples.

The Control dataset is made up of 60 real BBH merger events observed in LIGO and 60 glitches (10 of each kind) taken from the samples of glitches in the Test dataset.

For the MLGWSC-1 dataset, we use 6000 BBH samples generated using the technique mentioned in \cite{Schafer:2022dxv}. Since their study did not involve any glitches, we include the 6000 glitch samples from our test dataset. 

In Figure~\ref{fig:Network_ROCs}, we show the ROCs for these datasets. 
We see that the network performs the best on the Test dataset with an AUC score of 0.93. This is not surprising as the samples are taken from the same distribution as the training dataset. All the other curves, apart from the curves for the Control and MLGWSC-1 datasets, are close together at an average AUC score of 0.825. The two sets of curves - one comprising the Test, Control and MLGWSC-1 datasets and the other made up of the two Trigger dataset and the Hybrid dataset, differ only in the distribution of glitches. The set of curves with higher AUC scores have glitches selected at random from the entire GravitySpy Catalog and the other set of curves have glitches taken from the intersection between the GravitySpy and the other pipelines. 

The intersection between the samples identified in the \texttt{omicron} pipeline (GravitySpy is based on the triggers in the \texttt{omicron} Pipeline \citep{Zevin_2017}) and those in the other 4 pipelines is very small (as mentioned in section \ref{sec:data}), indicating fundamental differences in the sample pool and the sensitivities of the \texttt{omicron} pipeline compared to the other pipelines.
Our neural network seems to be sensitive to this change in the source pipeline, indicating that the network has picked up on the fundamental change between these pipelines, leading to a degradation in performance when tested on these samples.
This is further supported by the ROC of the Control dataset, which has only 60 real glitches but selected from the \texttt{omicron} pipeline similar to the glitches in our training dataset. Here, the performance is comparable to the Test dataset in spite of the sample size being limited.

Next, we analyse the MLGWSC-1 dataset, where the simulated BBH samples come from \cite{Schafer:2022dxv} but the glitches are the same as the Test dataset. The ROC curve of the network is closer to that of the Control dataset and Test dataset than the other curves. Therefore, the network retains its performance when we swap out simulated BBH samples for real BBH merger events or BBH merger events simulated and injected using a different approximant and technique. As long as the glitch samples are selected from the full \texttt{omicron} (GravitySpy) Catalog, the network performance is without any loss in accuracy.

We can also conclude that the network is very robust when it comes to classifying glitch classes never seen before. When tested on all glitch classes (including the \textit{None of the above}) despite being trained on only 6 classes of glitches the network was able to identify the new classes of glitches as glitches. This is evident when comparing the (Trigger dataset vs T$_{\rm{AG}}$dataset). On a separate note, comparing the Test dataset with Hybrid dataset shows a downgrade in the performance, and as argued before, it is due to the change in the primary search pipeline of the glitch samples. Therefore, the network performance is sensitive to the underlying search pipeline, and not the type of BBH signals (simulated or real) or the classes of glitches. In other words, there is a decline in the performance only when introduced to glitches identifed from a completely distinct search pipeline. Further investigations are needed to better understand the origins of differences in glitches found by different search pipelines.

\subsection{Performance on the real GW events}
Our goal is to gauge the performance of the network on the real BBH events, which we show as a function of their total masses and SNRs.

We test the network on three different sets, which may contain distinct subsets of real BBHs discovered through different search pipelines. 
The \texttt{omicron} dataset consists of O3 BBH merger events that were detected in the \texttt{omicron} pipeline. This dataset consists of 39 events. The O3 dataset consists of BBH events observed in the O3 run, regardless of the pipeline. The O3 dataset has 64 events. The \textit{All} runs dataset consists of the GW events from O1, O2, and O3 runs of the LIGO detector. It contains 75 events. These datasets do not contain all the events belonging to their respective category as we had some problems fetching the time series data for all the events.

Each dataset is split into 4 categories based on two thresholds - events from a system with a total mass above or below the  60~\Smass threshold, and events with an SNR above or below 12. 

We choose a threshold of 0.4 where class 0 is BBH and class 1 is glitch. For this threshold, the network has a F1 score of 0.86, an accuracy of 87\%, a TPR of 0.83 and a FPR of 0.1 on the Test dataset. 

Each cell in Figure \ref{fig:Real_GW_performance} shows the number of events classified correctly by the network for the selected threshold and the total number of events that belong to each category. The colours measure the accuracy of the network in each cell. Going from left to right, we see the performance of the network is relatively constant, apart from the effect of statistics. The category where the network seems to struggle the most in O3 dataset and All Runs Dataset is the high-mass-low-SNR category. This is expected as four out of the six types of glitches used in this work peak in the same region of the Sine-Gaussian space as the high-mass BBH events. This can be seen in Figure \ref{fig:ref_glitches}. But overall, the network can identify the BBH merger events correctly. Therefore, the network can sustain its performance when moving from simulated BBH merger signals to real BBH merger signals. 

\section{Summary and Conclusions} \label{sec:conclusions}
Rapid and accurate identification of GW signals continues to remain a significant challenge for the current and the future generations of GW detector observatories due to the ubiquitous presence of short-duration noise transients called glitches. Previously, in SiGMa-Net \citep{PhysRevD.107.024030}, we developed a simple convolutional neural network which made use of the Sine-Gaussian Projection (SGP) maps to distinguish (massive) BBHs from Blip glitches. 

In this work, SiGMa-Net II, we present an improved neural network to distinguish BBH signals from some of the most popular glitches found in O3 LIGO data, namely, Blips, Koifish, Tomte, Fast Scattering, Scattered Light, and Blip Low Frequency. The six types of glitches are selected from real data using the \texttt{omicron} pipeline and reported in the GravitySpy catalogue. Our BBH samples are simulated using the \texttt{PyCBC} and are injected into real noise, selected from relatively quiet data chunks in O3. We apply transfer learning and use the InceptionNetV3 architecture where the last few layers are trained on our own datasets. 

On the test dataset, our network gives a TPR of 0.83 and an FPR of 0.1 for a threshold of 0.4. Applying this as the optimal threshold on the sample of real BBH events, selected from observing runs O1 to O3, we essentially find that 80\% or more events are correctly identified. These events are distributed over a wide range of network SNRs ($6<\rm{SNR}<30$) and total masses ($14$M$_{\odot}<$~M$_{\rm{total}}<180$M$_{\odot}$). We note that the network efficiencies on the real and test datasets are consistent reflecting that our training-test datasets are likely representative and the network is not affected by overfitting issues. In other words, the network is robust to changes in detector and noise properties across the various runs of LIGO as shown by the retention of performance across different observing runs. 

Furthermore, the network achieves an accuracy of 82\% on the simulated BBHs of \cite{Schafer:2022dxv}, which is a completely independent dataset. We also find that even though the network is trained on six types of glitches from GravitySpy, it can still distinguish the BBH signals from other unseen types of glitches present in GravitySpy. However, the nature of triggers produced by the search pipelines like \texttt{GstLAL, MBTAOnline, PyCBC} and \texttt{SPIIR} seemed to have some fundamental differences compared to the glitches categorised by GravitySpy. The network performance declined in such cases as network training did not include these distinct kinds of triggers. Further investigation on the nature of these triggers and their categorisation, while important, is beyond the scope of this work.

With SiGMaNet II, we show the viability of using SGP maps and neural networks to efficiently distinguish the BBH from some of the most abundant glitch classes. In the near future, we plan to investigate the improvements arising from combining the complementary and more commonly used Q-Transform maps (time-frequency spectrograms) and dynamic combination of any two detectors within the network of GW observatories. We also would like to extend the analysis to even wider glitch classes, particularly, including those arising from the aforementioned standard search pipelines.   

We would be interested in studying the possible improvements that can come by including data from additional detectors such as the VIRGO and the KAGRA. Since more detectors would mean increased computation times not only to prepare input images but also to train the network, there will likely be a tradeoff between the pipeline speed and the gain achieved in the performance improvement. This study may also require exploration of more optimal architectures and data pre-processing protocols as a result.

\section*{Acknowledgements}
We would like to thank Siddharth Soni and Chayan Chatterjee for useful discussions. This material is based upon work supported by NSF's LIGO Laboratory which is a major facility fully funded by the National Science Foundation. We acknowledge the use of the IUCAA LDG cluster Sarathi for carrying out the computational and numerical work presented in this study. We acknowledge the use of the following \textsc{Python} packages: Numpy, Scipy, Pandas, Matplotlib, PyCBC.

\section*{Data Availability}
All training and testing data are available from the corresponding author upon reasonable request.

\bibliographystyle{mnras}
\bibliography{ref} 
\bsp
\label{lastpage}
\end{document}